\documentstyle[prl,aps,preprint,epsfig,placeins]{revtex}

\def\ifmath#1{\relax\ifmmode #1\else $#1$\fi}%
\def\GeV{\ifmmode \hbox{\rm\ Ge\kern -0.05em V}\else
                  \hbox{Ge\kern -0.05em V}\fi}%
\def\MeV{\ifmmode \hbox{\rm\ Me\kern -0.05em V}\else
                  \hbox{Me\kern -0.05em V}\fi}%
\def\keV{\ifmmode \hbox{\rm\ ke\kern -0.05em V}\else
                  \hbox{ke\kern -0.05em V}\fi}%
\def\eV{\ifmmode \hbox{\rm\ e\kern -0.05em V}\else
                 \hbox{e\kern -0.05em V}\fi}%

\newcommand{\pt}{\ifmath{p_T}}

\newcommand{\et}{\ifmath{E_T}}

\def\z0{\ifmath{Z^0}}%
\def\Z0{\ifmath{Z^0}}
\def\ee{\ifmath{e^+ e^-}}%
\def\mumu{\ifmath{\mu^+ \mu^-}}%

\def\zmumu{\ifmath{\z0 \rightarrow \mu^+ \mu^-}}
\def\zee{\ifmath{\z0 \rightarrow e^+e^-}}
\def\sumet{\ifmath{\sum \et^{jets}}}
\def\massbpr{\ifmath{m_{b'}}}

\def\btag{\ifmath{b}-tag}

\newcommand{\VPY}[3]{{\bf #1}, #2 (#3)}
\newcommand{\PRD}[3]{Phys. Rev. D \VPY{#1}{#2}{#3}}
\newcommand{\PRL}[3]{Phys. Rev. Lett. \VPY{#1}{#2}{#3}}
\newcommand{\PLB}[3]{Phys. Lett. B \VPY{#1}{#2}{#3}}
\newcommand{\NP}[3]{Nucl. Phys. \VPY{#1}{#2}{#3}}
\newcommand{\NIM}[3]{Nucl. Instrum. Methods A \VPY{#1}{#2}{#3}}

\def\etal{\emph{et al.}}
\newcommand{\fig}[1]{Figure~\ref{fig:#1}}
\newcommand{\tab}[1]{Table~\ref{tab:#1}}

\newcommand{\paper}[1]{~\cite{ref:#1}}

\begin{document}
\draft
\tightenlines

\title{\Large\bf Search for a Fourth-Generation Quark More Massive than
        the \z0\ Boson in $p\overline{p}$ Collisions at $\sqrt{s} = 1.8$~TeV}


\maketitle

\vspace{-3.2cm}
\rightline{FERMILAB-PUB-99/256-E}
\rightline{September 12, 1999}
\vspace{3.2cm}

\font\eightit=cmti8
\def\r#1{\ignorespaces $^{#1}$}
\hfilneg
\begin{sloppypar}
\noindent
T.~Affolder,\r {21} H.~Akimoto,\r {43}
A.~Akopian,\r {36} M.~G.~Albrow,\r {10} P.~Amaral,\r 7 S.~R.~Amendolia,\r {32} 
D.~Amidei,\r {24} J.~Antos,\r 1 
G.~Apollinari,\r {36} T.~Arisawa,\r {43}
% ...
\end{sloppypar}
\vskip .026in
\begin{center}
(CDF Collaboration)
\end{center}

\vskip .026in
\begin{center}
\r 1  {\eightit Institute of Physics, Academia Sinica, Taipei, Taiwan 11529, 
Republic of China} \\
\r 2  {\eightit Argonne National Laboratory, Argonne, Illinois 60439} \\
\r 3  {\eightit Istituto Nazionale di Fisica Nucleare, University of Bologna,
I-40127 Bologna, Italy} \\
\r 4  {\eightit Brandeis University, Waltham, Massachusetts 02254} \\
\r 5  {\eightit University of California at Los Angeles, Los 
Angeles, California  90024} \\  
\r 6  {\eightit Instituto de Fisica de Cantabria, University of Cantabria, 
39005 Santander, Spain} \\
\r 7  {\eightit Enrico Fermi Institute, University of Chicago, Chicago, 
Illinois 60637} \\
\r 8  {\eightit Joint Institute for Nuclear Research, RU-141980 Dubna, Russia}
\\
\r 9  {\eightit Duke University, Durham, North Carolina  27708} \\
\r {10}  {\eightit Fermi National Accelerator Laboratory, Batavia, Illinois 
60510} \\
\r {11} {\eightit University of Florida, Gainesville, Florida  32611} \\
\r {12} {\eightit Laboratori Nazionali di Frascati, Istituto Nazionale di Fisica
               Nucleare, I-00044 Frascati, Italy} \\
\r {13} {\eightit University of Geneva, CH-1211 Geneva 4, Switzerland} \\
\r {14} {\eightit Harvard University, Cambridge, Massachusetts 02138} \\
\r {15} {\eightit Hiroshima University, Higashi-Hiroshima 724, Japan} \\
\r {16} {\eightit University of Illinois, Urbana, Illinois 61801} \\
\r {17} {\eightit The Johns Hopkins University, Baltimore, Maryland 21218} \\
\r {18} {\eightit Institut f\"{u}r Experimentelle Kernphysik, 
Universit\"{a}t Karlsruhe, 76128 Karlsruhe, Germany} \\
\r {19} {\eightit Korean Hadron Collider Laboratory: Kyungpook National
University, Taegu 702-701; Seoul National University, Seoul 151-742; and
SungKyunKwan University, Suwon 440-746; Korea} \\
\r {20} {\eightit High Energy Accelerator Research Organization (KEK), Tsukuba, 
Ibaraki 305, Japan} \\
\r {21} {\eightit Ernest Orlando Lawrence Berkeley National Laboratory, 
Berkeley, California 94720} \\
\r {22} {\eightit Massachusetts Institute of Technology, Cambridge,
Massachusetts  02139} \\   
\r {23} {\eightit Institute of Particle Physics: McGill University, Montreal 
H3A 2T8; and University of Toronto, Toronto M5S 1A7; Canada} \\
\r {24} {\eightit University of Michigan, Ann Arbor, Michigan 48109} \\
\r {25} {\eightit Michigan State University, East Lansing, Michigan  48824} \\
\r {26} {\eightit University of New Mexico, Albuquerque, New Mexico 87131} \\
\r {27} {\eightit The Ohio State University, Columbus, Ohio  43210} \\
\r {28} {\eightit Osaka City University, Osaka 588, Japan} \\
\r {29} {\eightit University of Oxford, Oxford OX1 3RH, United Kingdom} \\
\r {30} {\eightit Universita di Padova, Istituto Nazionale di Fisica 
          Nucleare, Sezione di Padova, I-35131 Padova, Italy} \\
\r {31} {\eightit University of Pennsylvania, Philadelphia, 
        Pennsylvania 19104} \\   
\r {32} {\eightit Istituto Nazionale di Fisica Nucleare, University and Scuola
               Normale Superiore of Pisa, I-56100 Pisa, Italy} \\
\r {33} {\eightit University of Pittsburgh, Pittsburgh, Pennsylvania 15260} \\
\r {34} {\eightit Purdue University, West Lafayette, Indiana 47907} \\
\r {35} {\eightit University of Rochester, Rochester, New York 14627} \\
\r {36} {\eightit Rockefeller University, New York, New York 10021} \\
\r {37} {\eightit Rutgers University, Piscataway, New Jersey 08855} \\
\r {38} {\eightit Texas A\&M University, College Station, Texas 77843} \\
\r {39} {\eightit Texas Tech University, Lubbock, Texas 79409} \\
\r {40} {\eightit Istituto Nazionale di Fisica Nucleare, University of Trieste/
Udine, Italy} \\
\r {41} {\eightit University of Tsukuba, Tsukuba, Ibaraki 305, Japan} \\
\r {42} {\eightit Tufts University, Medford, Massachusetts 02155} \\
\r {43} {\eightit Waseda University, Tokyo 169, Japan} \\
\r {44} {\eightit University of Wisconsin, Madison, Wisconsin 53706} \\
\r {45} {\eightit Yale University, New Haven, Connecticut 06520} \\
\end{center}

\begin{abstract}

\hspace{\parindent}
We present the results of a search for pair production of a fourth-generation 
charge -1/3 quark ($b'$) in $\sqrt{s}=1.8$ TeV $p\overline{p}$ 
collisions using 88 pb$^{-1}$ of data obtained with the Collider Detector at 
Fermilab. We assume that both quarks decay via the flavor-changing neutral 
current process $b' \to b \z0$ and that the $b'$ mass is greater 
than $m_Z + m_b$.
We studied the decay mode $b'\overline{b'}\rightarrow \z0\z0 b\overline{b}$
where one \z0\ decays into
$e^{+}e^{-}$ or $\mu^{+}\mu^{-}$ and the other decays hadronically, giving
a signature of two leptons plus jets.
An upper limit on the
 $\sigma_{p\bar p \rightarrow b' \bar b'} 
\times \left[ BR \left(b' \rightarrow b\z0  \right)\right]^2$ is established 
as a function of the $b'$ mass.
We exclude at 95\% confidence level a $b'$ quark with mass between
100 and $199 \GeV/c^2$ for $BR(b' \to b \z0) = 100\% $. 

\end{abstract}



The Standard Model (SM) with three generations of quarks and leptons 
is in excellent agreement with all experimental data available today.
There is no strong reason to believe that an extra fermion generation 
exists. However, the SM does not explain either the fermion family 
replication or the fermion mass hierarchy.
Several models have been proposed to solve shortcomings in the
SM through the introduction of extra quarks and leptons, while
grand unified theories, 
supersymmetry, supergravity and
superstrings predict or can accomodate extra fermion 
states~\cite{ref:SM_extensions}. 
An extensive discussion of such models can be found  
in a recent review~\cite{ref:review}.


In general, flavor-changing neutral current (FCNC) processes 
in the Standard Model are highly suppressed. 
However, if a fourth-generation charge -1/3 
quark ($b'$) exists and is lighter than 
both the $t'$ (its partner in an SU(2) doublet) and the top quark ($t$), 
the charged-current (CC) decays $b' \to tW^{-}$ and $b' \to t'W^{-}$
are kinematically forbidden. The leading charged-current decay mode will 
then be $b' \to cW^{-}$, which is doubly Cabibbo-suppressed.
In this situation loop-induced FCNC decays can 
dominate~\cite{ref:review,ref:bprime_fcnc1,ref:bprime_fcnc2} provided
$|V_{cb'}|/|V_{tb'}|$ is less than roughly $10^{-2}$ to $10^{-3}$, 
depending on the $b'$ and $t'$ masses \cite{ref:bprime_fcnc2}.
If $m_{b'} > m_Z + m_b $, the dominant FCNC decay mode 
is $b' \to b \z0$~\cite{ref:bprime_fcnc2}
as long $b' \to b H$ is kinematically suppressed or 
forbidden~\cite{ref:bprime_higgs}.
For $m_{t} < \massbpr < m_{t}+m_{W}$, the decay mode $b' \to tW^{*}$ 
becomes available but is suppressed by three-body phase space, and
the $b' \to b \z0$ channel can still dominate over the CC decay for 
$b'$ masses up to about $230 \GeV/c^2$~\cite{ref:review,ref:bprime_tW*}.


Several experiments have searched explicitly for $b'$ quarks decaying via
FCNC~\cite{ref:experiments_FCNC}. 
The most stringent limit comes from the D\O\ Collaboration, which searched
in the  $b'\overline{b'} \to \gamma g b \overline{b}$ and
$b'\overline{b'} \to \gamma \gamma b \overline{b}$ channels, 
excluding a $b'$ quark mass up to $m_Z + m_b$
for a FCNC branching fraction larger than 50\%~\cite{ref:D0_limit}.
CDF has excluded a long-lived $b'$ quark with mass up to $148 \GeV/c^2$ and a 
lifetime of $\tau \approx 3.3 \times 10^{-11}$~sec, 
assuming $BR(b' \to b\z0) = 100\%$~\cite{ref:Peterson}.
If the CC decay $b' \to cW^-$ dominates, the lower mass bound of 128 GeV 
found in a D0 top quark search~\cite{ref:D0_top_search} also applies 
to the $b'$ quark~\cite{ref:PDG}. 

In this Letter, we report on a search for a $b'$ quark using 
$88 \pm 4$~pb$^{-1}$ of $p \overline{p}$ collisions at  $\sqrt{s} = 1.8$~TeV
collected with the CDF detector from 1994 to 1995.
Fourth-generation $b'$ quarks can be pair-produced in $p \overline{p}$ 
collisions through $gg$ fusion and
$q \overline{q}$ annihilation with the same cross section, for a given mass, 
as the top quark.
We search for pair-produced $b'$ quarks decaying via FCNC into $b\z0$,
where one \z0\ decays into leptons and the other decays hadronically.
The signature is two high transverse momentum (\pt ) 
leptons from the \z0\ decay, two
high-\pt\ jets from the second \z0, and two $b$ jets whose \pt\ scales with
the $b'$ mass.


A detailed description of the CDF detector can be found 
elsewhere\cite{ref:CDF_detector}. We briefly describe the components most 
relevant for this analysis.
Inside a 1.4 T solenoidal magnetic field, the silicon vertex detector (SVX), 
the  vertex time projection chamber (VTX), 
and the central tracking chamber (CTC) provide tracking information.
The SVX, positioned immediately outside the beampipe and inside the VTX,
consists of four layers of silicon micro-strip detectors and covers 
$|z| < 25$~cm~\cite{ref:coordinates}. 
It provides for precise track reconstruction in the plane transverse to the beam
and is used to identify secondary vertices from the decay of $b$ hadrons.
The VTX is used to measure the position of the primary 
interaction vertex along the $z$ axis.
The CTC is a cylindrical drift chamber 
that covers the pseudorapidity range $|\eta| < 1.1$ and 
consists of 84 layers that are grouped
in nine alternating superlayers of axial and stereo wires.
Outside of the solenoid, electromagnetic and hadronic calorimeters, arranged 
in a projective tower geometry, surround the tracking volume and are used to
identify electrons and jets over the range $|\eta| < 4.2$.
The electron energy is measured in the central electromagnetic 
calorimeter (CEM) ($|\eta| < 1.1$) and the end-plug electromagnetic calorimeter
(PEM) ($1.1 < |\eta| < 2.4$).
Outside the calorimeters, three systems of drift chambers in the 
region $|\eta|< 1.0 $ provide muon identification.


We select events satisfying a high-\pt\ lepton trigger, containing a 
well-identified muon or electron in the central region,
whose primary vertex is within 60 cm of the nominal interaction position.
A trigger that requires one jet with $\et > 10 \GeV$, in addition to
the lepton, is also used for muon events.
Inclusive \zee\ and \zmumu\ samples are selected by 
requiring one primary lepton that
satisfies tight lepton identification cuts and a second lepton satisfying
loose identification cuts~\cite{ref:thesis}.
Dielectron events are selected by requiring at least one tight electron 
with transverse energy $\et > 20 \GeV$ in the CEM and a second 
loose electron with $\et > 10 \GeV$ in either the CEM or PEM calorimeters.
Dimuon events are required to have one tight muon with 
transverse momentum $\pt > 20 \GeV/c$
in the central region and a second loose muon with $\pt > 10 \GeV/c$.
A calorimeter isolation cut is imposed on the second lepton.
We accept events if the reconstructed $ee$ or $\mu\mu$ invariant mass is 
between 75 and 105  GeV/$c^2$.
After this selection there are 6287 (2940) \z0\ events remaining in the electron
(muon) data sample.

\begin{figure}[bth]
\begin{center}
\epsfig{file=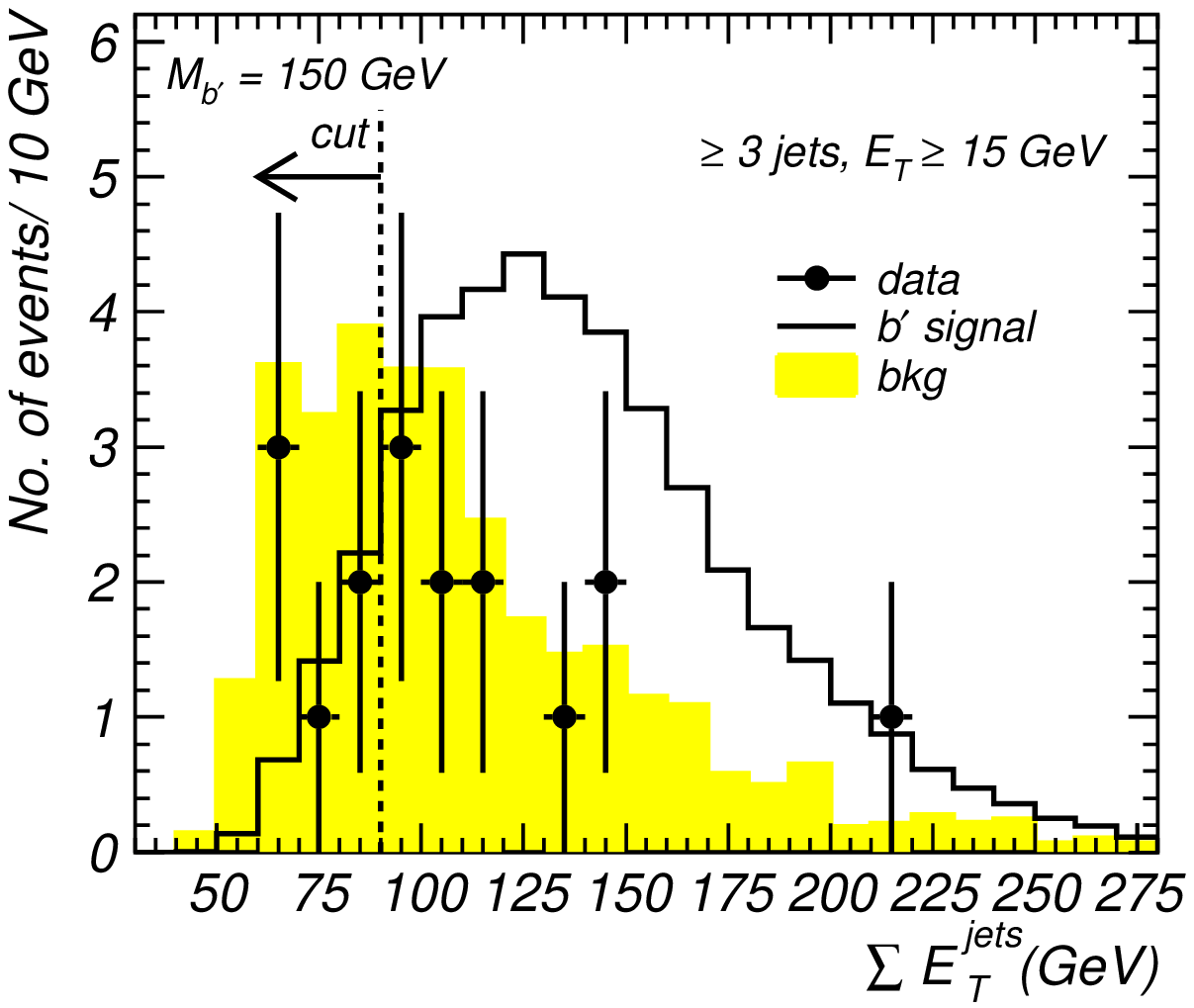,width=0.70\textwidth}
\end{center}
\caption{
\sumet\ distribution for events with at least 3 jets with
$\et > 15 \GeV$ and $|\eta| < 2$, before the \btag ging requirement.
The expected SM background is shown shaded.
The expected signal event distribution for a $b'$ quark mass of 150~GeV/$c^2$
is shown as a solid line.
The vertical dashed line represents the \sumet\ cut for this 
specific $b'$ mass. Events to the right of this line are accepted.}
\label{fig:sumet}
\end{figure}

In order to optimize our sensitivity to a  $b'$ quark signal
we make a jet selection that depends on the $b'$ mass being considered. 
Hadronic jets are selected using a clustering algorithm\paper{jets} 
with a cone size of $\Delta R = \sqrt{\Delta \eta^2 + \Delta \phi^2} = 0.4$.
Each event is required to have at least three jets within $|\eta| < 2.0$, 
two of which with $\et > 15 \GeV$.
For $b'$ masses above $120 \GeV/c^2$, the third jet is required to 
have $\et > 15 \GeV$. 
For $m_{b'} \leq 120 \GeV/c^2$, the \et\ requirement on the third jet
is relaxed to $\et > 7 \GeV$  since the $b$ jets for $b'$ masses near the 
$m_Z + m_b$ threshold have low momentum.
We define the variable \sumet\ as the 
summed transverse energy of jets with $\et > 15 \GeV$ and $|\eta| < 2.0$ 
and require this quantity 
to be larger than $\massbpr c^2 - 60 \GeV$.
\fig{sumet} shows the \sumet\ distribution for \ee\ and \mumu\ events passing 
the 3-jet requirement for $b'$ masses above $120 \GeV/c^2$.
Also shown are the distributions expected from SM background and from a
$b'$ quark with a mass of $150 \GeV/c^2$ (see below).


We further require at least one jet to be tagged as
a $b$ quark by the SVX \btag ging algorithm 
developed for the top quark analysis \cite{ref:secvtx}.
The number of events passing each major selection criterion for each
leptonic channel is shown in \tab{data}.
One $\mu\mu$ event passes all our selection criteria for  
$m_{b'} \leq 120 \GeV/c^2$.
This event has a third jet with $\et = 8.3 \GeV$ which fails the 
third-jet \et\ requirement for larger $b'$ masses.



The signal acceptance and detection efficiencies 
are estimated from a combination of
data and Monte Carlo simulation.
We have generated $b' \overline{b'} \to b\z0 \overline{b}\z0$ 
Monte Carlo samples for different $b'$ masses
between 100 and $210 \GeV/c^2$ using the HERWIG program\paper{herwig}
with MRSD0$^\prime$ structure functions~\cite{ref:MRSD0}.
One \z0\ is required to decay into muons or electrons while the other is allowed
to decay through any available decay channel.
The CLEO QQ Monte Carlo program \cite{ref:cleomc} is used to model the decays 
of $b$ hadrons. These events are passed through a simulation of the
CDF detector and subjected to the same selection requirements as the data.

The electron trigger efficiency is determined from data to be $(92 \pm 1) \%$, 
while the muon trigger efficiency per event $(82 \pm 4) \%$ is 
obtained from a combination of data and simulation.
The efficiencies of the lepton identification cuts are 
determined using a \zee\ (\mumu) data sample with an unbiased selection on one
of the leptons.
The \zee\ and \zmumu\ geometric and kinematic acceptance and detection 
efficiency, including the 
isolation efficiency, is $(41 \pm 3\%)$ and $(30 \pm 3)\%$
respectively and is nearly independent of the $b'$ mass.

The event \btag\ efficiency rises with $m_{b'}$ from 17\% for 
$m_{b'} = 100 \GeV/c^2$, to values between 50\% and 57\% for masses
above 150 GeV/$c^2$.
The total acceptance times efficiency, not including the $BR(Z \to l^+l^-)$, 
increases from 1.7\% (1.6\%) to 14\% (11\%)
for the electron (muon) channel as \massbpr\ increases 
from 100 to $210 \GeV/c^2$ (\tab{summary}).
This increase is due to the fact that a more massive
$b'$ leads to a more central event with more energetic jets 
in which, in addition, the \btag\ algorithm is more efficient.



The dominant systematic uncertainties on the acceptance times efficiency 
arise from the jet energy scale and gluon radiation~\cite{ref:thesis}.
By varying parameters in the Monte Carlo simulation
we estimate that the systematic uncertainty due to the jet energy scale 
in the electron (muon) channel is
16\% (14\%) for $m_{b'} = 100 \GeV/c^2$ and less than 13\% for higher masses. 
The presence of gluon radiation increases the jet multiplicity and therefore
increases the efficiency of the three-jet requirement. This effect is more
pronounced at low $b'$ mass because the $b$ quarks from low-mass $b'$ decay 
are produced near threshold and therefore are detected with low efficiency.
We estimate the systematic uncertainty due to this effect to be 19\% (18\%)
in the electron (muon) channel for $\massbpr = 100 \GeV/c^2$, and less than
9\% for a heavier $b'$.
Other important systematic uncertainties 
arise from the
\btag\ efficiency (10\%),
parton distribution function (5\%), 
total integrated luminosity (4.1\%), 
lepton identification efficiency (4\% for electrons, 5\% for muons), 
isolation efficiency (4\%) and 
trigger efficiency (1\% for electrons, 5\% for muons).
The total uncertainty on the acceptance times efficiency  
is shown as a function of $b'$ mass in \tab{summary}.


The only non-negligible background is from \z0\ events with associated 
QCD hadronic jets.
This background is estimated using a 
combination of the VECBOS \cite{ref:vecbos} and HERWIG Monte Carlo programs. 
VECBOS is used to make  the calculation
of the leading-order matrix elements for \z0\ + three partons 
events, with the MRSD0$'$ structure functions and the 
$\left < \pt \right >^2 $ of the generated partons for 
the QCD renormalization and factorization scales~\cite{ref:bkgpartons}.
A partial higher-order correction to the tree-level diagrams is obtained 
by including gluon radiation and hadronic fragmentation using HERWIG.
These \z0\ events are then passed through a simulation of the CDF detector.
We estimate the \btag\ rate in \z0\ plus jet events directly from data
using a technique developed for the top analysis
\paper{top_discovery}.
We apply the \btag\ rates measured in an inclusive jet sample
to the $\z0 +$~jets events that pass all the other 
selection criteria.
This method overestimates the background because the inclusive jet sample
contains heavy-quark contributions that are not present in \z0\ + jets events.
We expect approximately two background events for 
$\massbpr \leq 120 \GeV/c^2$ and less than one event for 
$\massbpr > 120 \GeV/c^2$,
in agreement with the number of events observed in the data.

\begin{figure}[btp]
\begin{center}
\epsfig{file=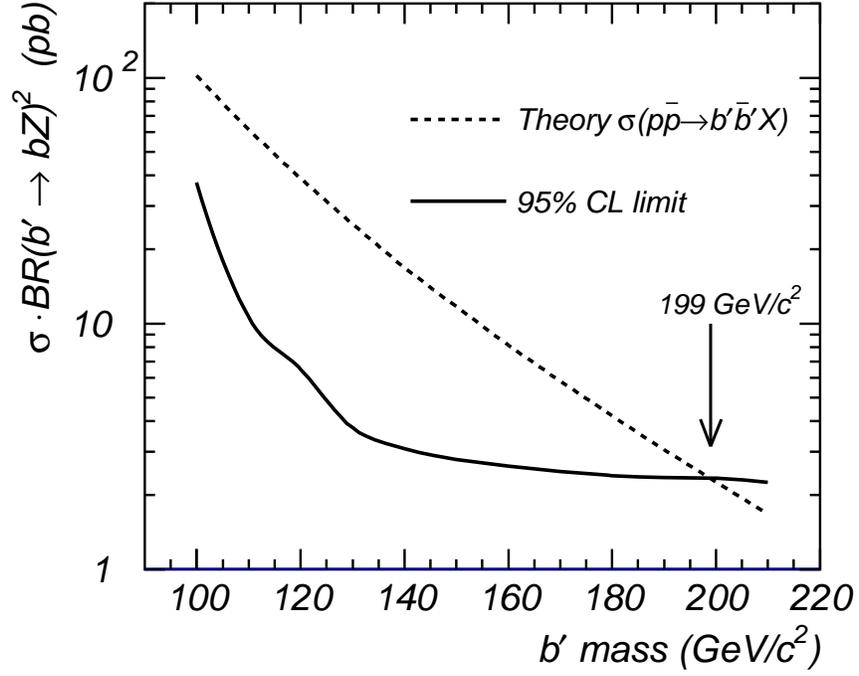,width=0.70\textwidth}
\end{center}
\caption{ The 95\% confidence level upper limit on 
$p \overline{p} \to b' \overline{b'}X$ production cross section times 
the $b' \to b\z0$ branching ratio squared (solid).
The dashed curve shows the predicted $\sigma_{p\bar p \rightarrow b' \bar b'} 
\times \left[ BR \left(b' \rightarrow b\z0  \right)\right]^2$ 
with the NLO production cross section from Ref. 23 
and $BR(b' \to b\z0 ) = 1$.}
\label{fig:limit}
\end{figure}

Under the assumption that the observed $\mu\mu$ event is from signal, 
that is, without subtracting background, we obtain a conservative 95\% 
confidence level upper limit on the
 $\sigma_{p\bar p \rightarrow b' \bar b'} \times 
\left[ BR \left(b' \rightarrow b\z0  \right)\right]^2$.
The limit is presented as a function of the $b'$ mass in \tab{summary}.
We have used a Bayesian method to calculate the limit and treat
the number of expected signal events as a Poisson distribution convoluted with 
a Gaussian systematic uncertainty.
Using the theoretical next-to-leading-order  $b'$ 
pair production cross section\paper{Laenen94} and assuming that
$BR(b' \to b\z0)$ is 100\%, we exclude at 95\% confidence level
$b'$ masses from $100 \GeV/c^2$ to $199 \GeV/c^2$, as shown in \fig{limit}.
This search is also sensitive to other $b'$ decay channels such as 
$b' \to bH$ or $b' \to {c}W^-$ as long $BR(b' \to bZ)$ is not
negligible, since the hadronic decays of the $H$ or $W$ are kinematically 
similar to those of the $Z$.
The acceptance for $b'\overline{b'} \to b \overline{b} ZH$ is 1.7
to 0.5 times the acceptance for $b'\overline{b'} \to b \overline{b} ZZ$,
depending on the Higgs and $b'$ masses and not including the
$BR(Z \to l^+l^-)$.
However, if we conservatively assume no sensitivity to 
these decay modes, we exclude a $b'$ mass from 104~GeV to 152~GeV for
$BR(b' \to bZ) \ge 50\%$.


     We thank the Fermilab staff and the technical staffs of the
participating institutions for their vital contributions.  This work was
supported by the U.S. Department of Energy and National Science Foundation;
the Italian Istituto Nazionale di Fisica Nucleare; the Ministry of Education,
Science, Sports and Culture of Japan; 
the Natural Sciences and Engineering Research Council of Canada; 
the National Science Council of the Republic of China; 
the Swiss National Science Foundation; the A. P. Sloan Foundation; and the
Bundesministerium fuer Bildung und Forschung, Germany.

\FloatBarrier



\begin{table}[bth]
  \begin{center}
    \caption{Events observed in data after each main selection requirement 
in both the electron and the muon channels.}
    {\footnotesize
        \begin{tabular}{ccccccc}
$m_{b'}$  & \multicolumn{3}{c}{\zee} & \multicolumn{3}{c}{\zmumu}  \\ 
\cline{2-4} \cline{5-7}
 (GeV/c$^2$) & 3 jets  & \sumet  & \btag  & 3 jets  & \sumet  & \btag \\ \hline
100 & 34 & 31 & 0 & 32 & 29 & 1 \\
120 & 34 & 20 & 0 & 32 & 21 & 1 \\
140 &  9 &  8 & 0 &  8 &  5 & 0 \\
160 &  9 &  4 & 0 &  8 &  4 & 0 \\
180 &  9 &  1 & 0 &  8 &  3 & 0 \\
200 &  9 &  1 & 0 &  8 &  2 & 0 
        \end{tabular}} 
\label{tab:data}
  \end{center}
\end{table}

\begin{table}[btp]
  \begin{center}
    \caption{Total acceptance ($A$) times efficiency ($\epsilon$) and 
relative systematic uncertainties ($\delta_{total}$) in the electron and 
muon channels, 95\% C.L. upper limit on the pair-production cross section 
times the branching ratio of $b' \to b\z0$ squared, and 
theoretical pair-production cross section~\protect\cite{ref:Laenen94}.}
    {\footnotesize
        \begin{tabular}{cdcdcdd}
         & \multicolumn{2}{c}{\zee} & \multicolumn{2}{c}{\zmumu} & & \\ 
\cline{2-3} \cline{4-5}
$m_{b'}$ & $(A \cdot \epsilon)$ & $\delta_{total}$ & $(A \cdot \epsilon)$ & 
$\delta_{total}$ & $\sigma \cdot BR^2_{95\%CL}$ & $\sigma_{theory}$\\
 (GeV/$c^2$) & (\%)& (\%) & (\%) & (\%) & (pb) & (pb)  \\ \hline
  100 & 1.7 & 29  &  1.6  & 27  &  37.  & 102.  \\     
  110 & 4.6 & 21  &  4.2  & 21  &  11.  & 61.6  \\ 
  120 & 7.6 & 20  &  6.4  & 19  &  6.5  & 38.9  \\ 
  130 & 8.2 & 19  &  6.8  & 19  &  3.8  & 25.4  \\ 
  140 & 9.9 & 19  &  8.3  & 19  &  3.1  & 16.9  \\ 
  150 & 11. & 19  &  9.2  & 19  &  2.8  & 11.7  \\ 
  160 & 12. & 19  &  9.8  & 19  &  2.6  & 8.16  \\ 
  170 & 12. & 19  &  10.  & 19  &  2.5  & 5.83  \\ 
  180 & 13. & 19  &  10.  & 19  &  2.4  & 4.21  \\ 
  190 & 13. & 19  &  11.  & 19  &  2.4  & 3.06  \\ 
  200 & 13. & 19  &  11.  & 19  &  2.4  & 2.26  \\ 
  210 & 14. & 19  &  11.  & 19  &  2.3  & 1.68  \\ 
        \end{tabular} }
\label{tab:summary}
  \end{center}
\end{table}


\begin{thebibliography}{99}
\bibitem{ref:coordinates} The CDF coordinate system uses $(\theta, \phi, z)$,
where the proton beam is in the $z$ direction and $\theta$ and
$\phi$ are the polar and azimuthal angles.
\bibitem{ref:bkgpartons} The outgoing partons are generated with a 
minimum \pt\ of $6 \GeV/c$, $| \eta | < 2.5$ and $\Delta R > 0.4$ 
between partons.
\end{thebibliography}
\end{document}
